\documentstyle[aps,preprint]{revtex} 

\begin{document}
{\flushright{IC-95-133}}

\begin{center}
\large {\bf A low $\alpha_s$: hint of new physics at the GUT scale?} \\
\vskip 1in
Mar Bastero-Gil$^{a}$ and Biswajoy Brahmachari$^{b}$ \\
\end{center}
\begin{center}
(a) Scuola Internazionale Superiore di Studi Avanzati \\ 34013
Trieste, ITALY. \\ (b) International Centre For Theoretical Physics,\\
34100 Trieste, ITALY.\\
\end{center}
\vskip 1in
{
\begin{center}
\underbar{Abstract} \\
\end{center}

In a SUSY GUT having an extra reverse doublet-triplet splitting near the 
GUT scale, where the mass of an extra doublet is greater than the mass of 
an extra triplet by two orders of magnitude, a low prediction of 
$\alpha_s$ can be achieved through threshold corrections via the
heavy scalars in the fundamental representations, making the 
prediction consistent with the values being suggested by low energy 
measurements. We implement this mechanism in a SUSY SU(5) GUT minimally 
extended to suppress Higgsino mediated proton decay. 
We also point out that as a consequence of this extension a natural 
scenario arises with a large hierarchy in Yukawa couplings ($ \lambda_t / 
\lambda_b \sim 40$). An experimental low 
value of $\alpha_s(m_Z)$ along with the non-observation of dimension-five 
proton decay modes at SuperKamiokande detector will favor such an extension 
of the SUSY SU(5) GUT over the minimal case.}


\section{Introduction}
The value of $\alpha_s$, calculated from a global fit of the LEP/SLC
data assuming Standard Model (SM) is correct, is three standard
deviations away from the value calculated from the low energy
experiments. Some time ago it was believed that the discrepancy is due
to the error in the low energy measurements. However, with the
increasing rigor of the lattice QCD calculations along with the wealth 
of low energy data, a real difference between the high and low energy 
predictions is emerging \cite{shifman}.  Even though the low energy 
experiments are indicating
a value of $\alpha_s=0.112$, the global fits at the Z-peak from the
LEP/SLC data assuming only the SM particle content and interactions
suggests a value of $\alpha_s \simeq 0.125$. Moreover, a high value of
$\alpha_s$ like 0.125 leads to a high value of $\Lambda_{QCD} \simeq$
500 MeV, differing from the measurement of the same from
the perturbative QCD and sum rules which is like 200 MeV.  Indeed the
work of Kane, Stuart and Wells \cite{kane} propose that a combined fit
of LEP/SLC data including the SUSY particles and interactions can
lower the value of $\alpha_s$. They have shown that to fit
$\alpha_s=0.112$ the $\chi^2$ minimization requires a chargino mass
near 80 GeV and a stop mass near 60 GeV. We note that, in stark
contrast, to predict a low value of $\alpha_s \sim 0.11$ in a minimal
SUSY SU(5) GUT, the mass scale of the superpartners has to be
considerably higher than the electroweak scale.
 
It is well known by now, that the coupling constant unification in
SUSY SU(5) model predicts a high value of $\alpha_s \sim 0.126~(0.130)$ for a
choice of the SUSY breaking scale $M_{SUSY}=1~TeV~(500~GeV)$. Light 
threshold effects tend to increase the
prediction of $\alpha_s$ even higher than the value obtained from LEP
data in the step function approximation or the so called run and match
method \cite{lang}. When an improved treatment of the low energy
threshold W
effects is done, including not only the leading logarithmic
contributions but also the finite part of the diagrams, the prediction
of $\alpha_s$ increases even further
\cite{mar,bagger}. At the same time when the heavy spectrum is 
non-degenerate, the heavy threshold correction 
to the prediction of $\alpha_s$ comes from the split incomplete SU(5)
multiplet containing the color triplet Higgs field \cite{hallsarid}.
This correction has an increasing effect on $\alpha_s$ whenever the
mass of the color triplet is greater than the mass of the doublet. It
has been noted
\cite{pnath,hisano,yamada} that there exist stringent lower bounds on
the mass of the color triplet coming from Higgsino mediated proton
decay and hence in the minimal model the doublet-triplet splitting
sizably increases the predicted value of $\alpha_s$ again.

There has been a number of attempts to lower the prediction of
$\alpha_s$ in a supersymmetric GUT. In the low energy scales the Winos
and the Binos give a threshold correction to $\alpha_s$ which has an
opposite sign to the threshold correction induced by gluinos;
consequently, it has been pointed out by Shifman and Roszkowski
\cite{rosz} that if one gives up the unification of the gaugino masses
a low prediction of $\alpha_s$ can be obtained due to a large
correction coming from the mass difference between the charginos and
the gluinos. On the other hand by invoking an intermediate B-L
symmetry breaking scale can also get a low prediction of $\alpha_s$.
Such a scenario has been explored in Ref \cite{moha} by introducing an
enlarged scalar sector inspired by superstring theory. Another
possibility is to introduce  higher dimensional SU(5) Higgs
multiplets. If the $50$, $\overline{50}$ and $75$ dimensional Higgs
fields of SU(5) are introduced \cite{masiero}, the heavy threshold
effects coming from the mass splitting within these extra multiplets
can also lower the prediction of $\alpha_s$ \cite{yamada}.

In this paper we stick to the conventional one step breaking of a SUSY
GUT model without giving up the universality of the gaugino masses 
\cite{nilles} at
the unification scale and consider the possibility of achieving a low
$\alpha_s$. We explore a possible reverse doublet-triplet splitting
which will have an effect opposite to the conventional doublet-triplet
splitting on the prediction of $\alpha_s$. Such a strange reverse
doublet-triplet splitting is indeed possible in a realistic SU(5)
model as will be displayed in this letter.

This paper is organized as follows. In section II we give the mechanism, 
in section III we implement it in a model, in section IV we discuss $m_t 
/ m_b$ in this new model, in section V we note some observations 
regarding the model and in section VI we conclude.

\section{Mechanism}
At first let us consider the prediction of $\alpha_s$
including the threshold effects in SUSY SU(5), which is well-studied
in the literature 
\cite{mar,bagger,hisano,yamada,barhall}. Throughout
this paper we will assume that including the threshold corrections,
the minimal SUSY SU(5) GUT predicts $\alpha_s=0.126$ \cite{bagger}; we
will also assume that the mass of the color triplet Higgs scalars in a
minimal SU(5) GUT is $10^{16.6}$ GeV\footnote{ This stringent
lower bound comes from the non-observation of the dimension five proton
decay processes assuming $M_{SUSY} \le O(TeV)$\cite{mar,hisano}. We are 
taking the lowest allowed value of $M_3$ which leads to 
the lowest $\alpha_s(m_Z)$.}. In particular the 
prediction of $ \alpha_s$ in the minimal SUSY SU(5) can be written as,
\begin{equation}
\alpha^{-1}_s(m_Z)={1 \over 2} ~[3 \alpha^{-1}_2(m_Z)-\alpha^{-1}_1(m_Z)] 
- { 3 \over 5 \pi} \ln[{M_3 \over M_2}] + T_L + \delta_{2loop}
\label{a3su5} ,
\end{equation}
where, $M_3$ and $M_2$ are the masses of the triplet and the doublet
Higgs scalars present in the $5$ and $\overline{5}$ representations of
SU(5), and $T_L$ parametrizes the contribution from all other light
degrees of freedom (excluding the light Higgs doublets),
and in a simple step function approximation \cite{hisano} $T_L= {1 \over 
2\pi} \ln {M_{SUSY} \over m_Z}$. $M_{SUSY}$ can be considered in the 
simplest
approach as a common SUSY breaking scale, or as an effective SUSY mass
parameter \cite{lang} resumming the effect of the detailed SUSY
spectrum, and in this sense it can be either more or less than $m_Z$
depending on the super-partner masses.  Here we assume that the
supersymmetric particle masses are not much higher than 1 TeV and
hence, the term $T_L$ is not enough to lower the prediction of
$\alpha_s$ consistent with the low energy experiments. The term 
$\delta_{2loop}$ parametrizes the two loop corrections (due to the light as 
well as the heavy degrees of freedom). In a 
generic situation one has $M_3>M_2$ and consequently the doublet-triplet
splitting increases the prediction of $\alpha_s$ via the second term
of Eqn (\ref{a3su5}). However, we notice the hypothetical possibility,
that if the mass of the doublet were more than the mass of the
triplet, we would have had a reverse effect on $\alpha_s$.  Keeping
this in mind we add one more $5+\overline{5}$ Higgs scalars with
doublet and triplet masses as $M^\prime_2$ and $M^\prime_3$ GeV
respectively. In that case the Eqn (\ref{a3su5}) gets modified to,
\begin{equation}
{\alpha^\prime}^{-1}_s(M_Z)={1 \over 2} ~[3
\alpha^{-1}_2(m_Z)-\alpha^{-1}_1(m_Z)] - { 3 \over 5
\pi}
\ln[{M_3 M^\prime_3
\over M_2 M^\prime_2}] + T_L + \delta^{\prime}_{2loop} \label{a3}.
\end{equation}
Taking the difference of Eqn (\ref{a3su5}) and Eqn (\ref{a3}) and
assuming,
\begin{equation}
M_3=10^{16.6}~;~M_2=10^2~;~M^\prime_3=10^x~;~M^\prime_2=10^y,
\end{equation}
we get,
\begin{equation}
\Delta \alpha^{-1}_s \equiv {\alpha^\prime}^{-1}_s(m_Z)-\alpha^{-1}_s(m_Z) 
={3 \over 5 \pi} (y-x) \ln 10+ [\delta_{2loop} -
\delta^\prime_{2loop}].
\label{diff} 
\end{equation}
It is easy to check from Eqn (\ref{diff}) that taking $y-x=2.26$ we
can get $\Delta \alpha^{-1}_s =0.99$ and consequently $\alpha_s$
decreases by 11\%, from 0.126 to 0.112. We have neglected the
$difference$ in the 2 loop terms which is due to the difference between
the two heavy sectors only. Instead if we add n extra pairs of
$5+\overline{5}$ the required splitting in each SU(5) multiplet is
only 2.26/n orders of magnitude.

It is important to check the change in the gauge boson masses due to
this extra splitting, because, as a general trend [see the formula
below] a reduction in the predicted value of $\alpha_s$ is associated
with a reduction in $M_V$ which is the mass of the heavy gauge bosons
mediating the dimension six proton decay processes. In the minimal
SUSY SU(5) GUT the the mass of the heavy gauge bosons can be predicted
from the following formula,
\begin{equation}
\ln{M_V \over m_Z}= { \pi \over 2} [\alpha^{-1}_1(m_Z)-\alpha^{-1}_2(m_Z)]+ 
{1 \over 10} \ln[ {M_3
\over M_2}] - { 1 \over 2} \ln[{M_\Sigma \over m_Z}] + T^1_L + 
\Delta_{2loop}, \label{mvsu5} \end{equation} 
where, $T^1_L$ parametrizes the threshold effects coming from the
fields present in the low energy scales, and $M{_\Sigma}$ is the mass
of the heavy Higgs scalar in the adjoint representation. In a simple
step function approximation, $T^1_L=-{5 \over 12} \ln{ M_{SUSY} \over
m_Z}$. At this stage we introduce the extra $5+\overline{5}$ multiplets. 
Now Eqn (\ref{mvsu5}) looks as,
\begin{equation}                                                              
\ln{M^\prime_V \over m_Z}= { \pi \over 2} [\alpha^{-1}_1(m_Z)-\alpha^{-1}_2(m_Z)]+ { 1 \over 10} \ln[{ 
M_3 M^\prime_3 \over M_2 M^\prime_2}] - { 1 \over 2}
\ln[{M_\Sigma \over m_Z}] +  T^1_L + \Delta^\prime_{2loop}. \label{mvnew}
\end{equation} 

We compare with minimal SUSY SU(5) case as before and obtain,
\begin{equation}
\ln[{M^\prime_V \over M_V}] = - { 1\over 10} (y-x) \ln 10 + 
[\Delta_{2loop} - \Delta^\prime_{2loop}]\label{mprime}. 
\end{equation}
Using $y-x=2.26$ and neglecting $\Delta_{2loop} -
\Delta^\prime_{2loop}$, we have, 
\begin{equation}
{M^\prime_V \over M_V}=10^{-0.226}.
\end{equation}

Clearly, this is a small reduction, and is consistent with the bounds on
$M^\prime_V$ from dimension six proton decay. In practice,
Eqn (\ref{mvsu5}) and Eqn (\ref{mvnew}) only determines the combination
${{M^\prime_V}^2 M_\Sigma \over m^3_Z}$. The change in $M_V$ can in
principle be compensated by a suitable lowering of $M_\Sigma$ too.  On
the other hand, $M_\Sigma$ is determined in terms of $M^\prime_V$
modulo the unknown Yukawa coupling of the $24^3$ term in the
superpotential. In a natural scenario $M_\Sigma$ is not expected to be
much lower than $M^\prime_V$. In the present case, the smallness of
the change in $M_V$ due to the introduction of the extra
$5+\overline{5}$ scalars assures that the fine-tuning of the mass
$M_\Sigma$ is not necessary.

If the extra triplet couples to the fermions it leads to dimension-five 
proton decay diagrams and consequently
its mass is bounded from below to $10^{16.6}$ GeV. 
In such a situation
the extra doublet has to have a mass at the Plank scale to lower the
prediction of $\alpha_s$. 
This rigid situation can be relaxed if we
introduce a mechanism to
suppress the Higgsino mediated proton decay in a SU(5) theory so that the 
mass of the
triplet can safely be lowered below the $10^{16.6}$ GeV scale. We will 
discover below, that the mechanism to suppress Higgsino mediated proton decay
naturally introduces extra $5+ \overline{5}$ Higgs scalars also.
\section{Model}
Babu and Barr \cite{babr} have shown that it is possible to suppress
the Higgsino mediated proton decay strongly in an SO(10) model by a
judicious choice of the fields, couplings and VEVs at the GUT scale.
Here, we consider a similar scenario in a SUSY SU(5) GUT.
Consider an SU(5) invariant superpotential involving the three pairs
of scalar superfields $\overline{5_i}$ and $5_i$ in the fundamental
representation, the $\Sigma$ superfield in the adjoint
representation, and three singlet superfields $\eta_1$, $\eta_2$ and 
$\eta_3$. The
fermions are in the usual $\overline{5_F}$ and $10_F$ representations
of the SU(5) group. The superpotential is,
\begin{eqnarray}
W &=& \overline{5_1} (M_{12}+\lambda_{12}~\Sigma)  5_2 
+\overline{5_2} (M_{21}+\lambda_{21}~\Sigma)  5_1 \nonumber\\
&+&
\overline{5_2} \lambda_{23}~\eta_1  5_3 +
\overline{5_3} \lambda_{32}~\eta_1  5_2 +
\overline{5_3} \lambda_{33}~\eta_2  5_3 
\nonumber\\ 
&+&M~\Sigma^2 + \beta~\Sigma^3 \nonumber\\
&+&\lambda^\eta_{1}~\eta^2_2\eta_3 + 
+M^\eta_{13}~\eta_1\eta_3 
\nonumber\\
&+&\lambda_d ~\overline{5_1}~\overline{5_F}~10_F +
\lambda_u~5_1~10_F~10_F.
\label{potential}
\end{eqnarray}
The $Z_N$ symmetry which allows only these couplings and forbids the
rest of the SU(5) invariant couplings in the scalar sector is given in
Table \ref{table1}.


We notice that only $5_1$ and $\overline{5_1}$ fields can couple to
the fermions and they do not have a direct mass term. This suppresses
the Higgsino mediated proton decay.  The scalars $\Sigma$, $\eta_1$, 
$\eta_2$ get VEVs at the GUT scale whereas $\eta_3$ does not get a VEV.  
The singlet $\eta_3$ couples to $\eta_1$ and $\eta_2$ only and it is 
required to
guarantee a consistent set of minimization conditions. The form of the
mass matrix can be easily calculated from the superpotential in Eqn 
(\ref{potential}). The general form of the mass matrices of the triplet
($M^T$) and the doublet($M^D$) is given by,
\begin{equation}
M^T=\pmatrix{ 0&a_3&0 \cr b_3&0&c \cr 0&d & e}~~,
~~M^D=
\pmatrix{ 0&a_2&0 \cr b_2&0&c \cr 0&d&e}
\end{equation}
To generate one vanishing eigenvalue we need to keep the usual form of 
the fine-tuning of the minimal SU(5) model
in the doublet sector to the relation,
\begin{equation}
b_2=M_{21}-3 \lambda_{21} v=0, \label{finetune}
\end{equation}
where we have used,
\begin{equation}
\langle \Sigma \rangle = diag~(2v,2v,2v,-3v,-3v).
\end{equation}
All these eigenvalues receive
corrections of the order of $m_{3/2}$ due to the soft breaking of
supersymmetry. Now the model has three heavy triplets, two doublets
having masses $d_1$ and $d_2$ at the unification scale and one doublet of mass of
order $m_{3/2}$. Following the definition in Eqn (\ref{diff}), and
denoting the triplet-mass\footnote{The masses of the weak scale doublets 
cancel out in the ratio.} in the minimal SU(5) case as $M^0_3$, 
\begin{equation}
\Delta \alpha^{-1}_s = 
{3 \over 5 \pi} \ln |{ M^0_3 d_1 d_2 \over {\rm Det}~(M^T)}|=
{3 \over 5 \pi} \ln |{ M^0_3 \sqrt{a^2_2 c^2+c^2 d^2 + e^2 a^2_2} 
\over a_3 b_3 e}|. 
\label{result} 
\end{equation}
\section{Top and bottom quark mass splitting}
At this point we note that
in this model there is an interesting possibility to generate
a hierarchy of Yukawa couplings to the fermions in the up and in
the down sector. When the SU(5) symmetry is broken there is one
combination of the $5_i$ fields and one combination of $\overline{5}_i$ 
fields remaining light. Denoting them with a superscript zero we can 
explicitly write down the combinations as,
\begin{eqnarray}
{5_1}^0 &=& 5_1, \nonumber\\
{\overline{5_1}}^0 &=& 
\gamma_1\overline{5_1}+\gamma_2\overline{5_2}+
\gamma_3\overline{5_3}.
\end{eqnarray}
It is straightforward to calculate the coefficients $\gamma_i$ from the 
doublet mass matrix. In particular,
\begin{equation}
\gamma_1=
{1 / \sqrt{1+{ a^2_2 \over d^2}+ {a^2_2 e^2 \over d^2 c^2}}}.
\end{equation}
Noting that at the GUT scale only the $5_1$ and 
$\overline{5_1}$ couple to the fermions, we can write down the
effective Yukawa couplings of fermions to the ${\overline{5_1}}^0$ and 
${5_1}^0$ scalars, as, 
\begin{equation}
\lambda^{eff}_u= \lambda_u~~~;~~~\lambda^{eff}_d= \gamma_1\lambda_d. 
\end{equation}
The doublets residing in the ${\overline{5_1}}^0$ and ${5_1}^0$ gets
the weak scale VEV leading to a splitting in the masses of the
top quark and the bottom quark. In the 
case of $\lambda_u=\lambda_d$ as well as small $\tan \beta$ we have to have,
\begin{equation}
\gamma_1 \sim {m_b \over m_t} \sim (2.3-2.5) \times 10^{-2}, \label{tb} 
\end{equation}
using $m_t=176$ GeV and $m_b=4.1-4.5$ GeV.
Now we are in a position to analyze 
the parameter space of Eqn. 
(\ref{result}) using Eqn. (\ref{tb}) as that of a constraint. Expressing 
Eqn. (\ref{result}) in terms of $\gamma_1$ we obtain,
\begin{equation}                                                                
\Delta \alpha^{-1}_s =                                                    
{3 \over 5 \pi} \ln |{ M^0_3  \over b_3} {c d \over a_3 e} 
\gamma^{-1}_1|.             
\label{result1}                                                            
\end{equation}                                                                  
It is not difficult to lower the prediction of $\alpha_s(m_Z)$ satisfying 
Eqn. (\ref{tb}); as an example,
\begin{equation}
{a_3 \over d} \sim 0.63~~;~~{a_2 \over d} \sim 6.31~~;~~{e \over c} \sim 
6.31~~;~~{M^0_3 \over b_3} \sim 10, \end{equation} we can achieve $\Delta 
\alpha^{-1}_s =0.88$ and $\gamma_1 = .025$. This lowers the prediction of 
$\alpha_s(m_Z)$ to 0.113 from 0.126. This is the lowest value we could 
achive keeping all the mass ratios less or equal to 10 $and$ $\gamma_1 = 
{m_b \over m_t}$. We note that this is not an unique prediction, and 
$\alpha_s(m_Z)$ could be larger or smaller as well, simply 
because the threshold corrections depend on the heavy masses of the 
model. All we are pointing out is that due to the reverse 
doublet-triplet splitting there exists a range of parameter space 
which can accomodate the lower values of $\alpha_s(m_Z)$ unlike the 
minimal SU(5) case.

\section{Observations}
Before we conclude we observe the following points.

(1) In the minimal case, to achieve a prediction of $\alpha_s=0.126$ one 
needs a SUSY spectrum\footnote{This can be understood from the 
form of the term $T_L$ in Eqn (\ref{a3su5}). For 
rigorous details see Ref. \cite{lang,mar,bagger}.} of the order of 1 TeV or 
more. However, in the present case, due to the influence of the extra heavy 
threshold effects, the SUSY spectrum can be as low as the weak scale, and 
still the prediction of $\alpha_s$ can be kept under desirable control. 
This makes the present case interesting for the future collider searches.

(2) In a SO(10) scenario as discussed by Babu and Barr 
\cite{babr} the
Higgs scalars in the adjoint representations 45 and 54 of SO(10) lead
to the threshold corrections to $\alpha_s (m_Z)$.  On the contrary,
splitting in the adjoint 24 of SU(5) has no threshold correction to
the low energy prediction of $\alpha_s(m_Z)$\footnote{ The reason 
lies in the special group theoretical decomposition of the 24 scalar under 
the low energy group. After decomposition, the individual contributions 
cancel each other in Eqn (\ref{a3su5}).}. Thus, in a SU(5) theory, the 
simplest way to alter the prediction of $\alpha_s(m_Z)$ is via the 
splitting in the extra 
$5+\overline{5}$ scalars. Interestingly, the scenario presented above, needs 
only singlets, fundamentals and adjoint of SU(5), and hence, one is not 
forced to introduce scalars in the higher dimensional tensorial 
representations of SU(5) \cite{bagger,yamada} to lower the prediction of 
$\alpha^{-1}_s(m_Z)$.

(3) We digress for a while to the question of R-parity 
violation \cite{rpty}. In the minimal SU(5) model the matter-parity 
violating coupling $\overline{5_F}~\overline{5_F}~10_F$ is allowed in the 
superpotential at the renormalizable level. This coupling leads to the 
unsuppressed baryon and lepton number violations, which in turn give 
rise to catastrophic proton decay unless the strength of this coupling 
is assumed to be vanishing by fiat.  In the present model, the $Z_9$ 
symmetry forbids this dangerous matter parity violating term offering a 
natural explanation of R-parity conservation at low energy.
\section{Conclusion}
To conclude, we have briefly reviewed the fact that in the minimal
SU(5) GUT the $prediction$ of $\alpha_s(m_Z)$ is at least as high as 0.126,
which is inconsistent with the $extraction$ of the same from the low
energy measurements that give a value around 0.112. Following the 
construction of an explicit SU(5) model, where the Higgsino mediated proton 
decay is strongly suppressed \cite{babr}, the heavy threshold corrections 
to the prediction of $\alpha_s(m_Z)$ have been calculated. Such a 
model naturally calls for the introduction of extra $5+\overline{5}$ 
Higgs scalars. Small mass splittings between the doublets and the 
triplets residing in the extra $5+\overline{5}$ representations lead to GUT 
scale threshold corrections to lower the prediction of $\alpha_s(m_Z)$ 
making it consistent with the values being extracted from the low energy
measurements. This model has a natural scenario to explain the 
large ratio of Yukawa couplings $\lambda_t / \lambda_b \sim 40$. This study 
shows that if the experimentally extracted 
value of $\alpha_s(m_Z)$ turns out to be lower than 0.126, and at the 
same time the dimension five proton decay modes are not observed in 
SuperKamiokande detector, it will favor the extended SU(5) theory 
described above over the minimal SU(5) theory. 

We acknowledge insightful communications with K. S. Babu, R. N. Mohapatra 
and J. P\'erez--Mercader.

\begin{table}[]
\begin{center}
\[
\begin{array}{|c||c||c||c||c||c||c||c||c||c||c||c|}
\hline
\overline{5_1}& 5_1& \overline{5_2}& 5_2&\overline{5_3}&
5_3&\Sigma&\eta_1&\eta_2&\eta_3 & \overline{5_F}& 10_F \\ \hline   
x^3&1&1&x^6&x^7&x^4&1&x^5&x^7&x^4&x^6&1\\
\hline
\end{array}
\]
\end{center}
\caption{The $Z_9$ charges assigned to various superfields. In our
notation $x^9=1$} \label{table1}
\end{table} 

\end{document}